\documentclass{PoS}
\usepackage{amsmath}
\usepackage{slashed}
\usepackage{subfigure}

\newcommand{\beq}{\begin{equation}}
\newcommand{\eeq}{\end{equation}}

\title{Renormalisation constants of quark bilinears in lattice QCD with four 
dynamical Wilson quarks}

\ShortTitle{Quark Bilinear RCs with $N_f=4$ dynamical flavours}

\author{ETM Collaboration}
\author{Benoit~Blossier$^{a}$, Mariane~Brinet$^{b}$,
Nuria~Carrasco$^c$, Petros~Dimopoulos$^d$, 
Xining~Du$^{b}$,
Roberto~Frezzotti$^{d,e}$, Vicent~Gimenez$^{c}$, 
        Gregorio~Herdoiza$^{f}$, Karl~Jansen$^{g}$, Vittorio~Lubicz$^{h}$,
         \speaker{David~Palao}$^e$, Elisabetta~Pallante$^{i}$,  Olivier~P\`ene$^{a}$,
          Konstantin~Petrov$^{a}$,   
          Siebren~Reker$^{i}$,   Giancarlo~Rossi$^{d,e}$, 
          Francesco~Sanfilippo$^{a}$,    
          Silvano~Simula$^{j}$, Luigi~Scorzato$^{k}$, Carsten~Urbach$^{l}$\\
        \llap{$^a$}  Laboratoire de Physique Th\'eorique, 
          CNRS et Universit\'e Paris-Sud XI, \\ B\^atiment 210, 91405 Orsay Cedex, France\\  
          \llap{$^b$} Laboratoire de Physique Subatomique et de Cosmologie, 
         CNRS/IN2P3/UJF, 53 avenue des Martyrs, 38026 Grenoble, France\\
        \llap{$^c$}Dep. de Fisica Te\'orica and IFC, Universitat de Valencia-CSIC,\\ 
        Dr. Moliner 50, E-46100 Burjassot, Spain \\   
        \llap{$^d$}Dipartimento di Fisica, Universit\`a di Roma ``Tor Vergata''\\
        Via della Ricerca Scientifica 1, I-00133 Rome, Italy\\
        \llap{$^e$}INFN Sezione di ``Roma Tor Vergata'', Via della 
                   Ricerca Scientifica 1, I-00133 Rome, Italy\\
        \llap{$^f$}Dep. de F\'{\i}sica Te\'orica and Instituto de F\'{\i}sica Te\'orica UAM/CSIC, \\
           Universidad Auton\'oma de Madrid, E-28049 Cantoblanco, Spain\\
        \llap{$^g$}DESY\footnote{Preprint number: DESY 11-235}, Platanenallee 6, D-15738 Zeuthen, Germany\\  
        \llap{$^h$}Dipartimento di Fisica, Universit\`a  Roma Tre and INFN\\
        Via della Vasca Navale 84, I-00146 Rome, Italy\\ 
        \llap{$^i$} Centre for Theoretical Physics, University of Groningen, \\ Nijenborgh 4,  
             9747 AG Groningen, The Netherlands \\ 
       \llap{$^j$} INFN-Roma Tre Via della Vasca Navale 84, I-00146 Rome, Italy\\  
         \llap{$^k$} ECT*/FBK -  strada delle tabarelle 286 - 38123 - Villazzano (TN) - Italy\\ 
        \llap{$^l$}  HISKP (Theory) and  Bethe Center for Theoretical Physics, 
                Rheinische Friedrich-Wilhelms-Universit{\"a}t Bonn,
                 Nussalle 14-16, 53115 Bonn, Germany  \\       
        E-mail: \email{david.palao@roma2.infn.it}}

\abstract{We present preliminary results of the non-perturbative computation of the RI-MOM 
renormalization constants in a mass-independent scheme  for the action with Iwasaki
glue and four dynamical Wilson quarks employed by ETMC. Our project
requires dedicated gauge ensembles with four degenerate sea
quark flavours at three lattice spacings and at several values of the standard and twisted quark mass
parameters. The RI-MOM renormalization constants
are obtained from appropriate $O(a)$ improved estimators extrapolated to
the chiral limit.  }

\FullConference{ The XXIX International Symposium on Lattice Field Theory - Lattice 2011\\
July 10-16, 2011\\
Squaw Valley, Lake Tahoe, California}

\begin{document}

\section{Introductory remarks and computational setup}
The European Twisted Mass Collaboration (ETMC) has recently performed simulations 
employing mass degenerate light (up/down) doublet quarks and a heavy mass 
non-degenerate pair for strange and charm quarks entering in the 
era of precise and realistic $N_f=2+1+1$ dynamical quark lattice computations~\cite{ETMC2p1p1}.   
In the ETMC setup gluon interactions are described by the Iwasaki action~\cite{Iwasaki}. Fermions are 
regularised in the maximally twisted mass (Mtm) Wilson 
lattice formulation~\cite{FR1,FRnondegproc,FR2}. 
This choice of the fermionic action has the benefit of achieving automatic O$(a)$-improvement, generally  
leading to small O$(a^2)$ lattice artefacts. Up to now ETMC has produced $N_f=2+1+1$ 
dynamical quark gauge configurations at three values of the lattice spacing 
(namely $a\sim 0.06, 0.08$ and 0.09 fm). 
Gauge ensembles have been produced at several values of the quark masses with 
the lowest pseudoscalar mass being of about 270 MeV. Obviously the 
inclusion of the dynamical strange and charm quarks offers the advantage of taking into account a rather 
important source of systematic effects. 

Computation of renormalisation constants (RCs) is 
a crucial step in order to extract
physical quantities from 
lattice data. It is worth noticing that the Mtm setup offers a rather convenient 
quark mass renormalisation pattern. Indeed, for the renormalisation of the masses of degenerate and 
non-degenerate quark pairs one only needs to know the non-singlet $Z_P$ and $Z_S$ 
renormalisation constants (and not \(Z_{S^0}\))~\cite{FRnondegproc}. 
RCs of operators with non-zero anomalous
dimension need to be computed in the 
chiral limit and for this reason dedicated lattice simulations employing $N_f=4$ 
light and (for simplicity) degenerate dynamical quarks are required.
Indeed we have produced $N_f=4$ gauge configuration ensembles 
corresponding to several sea quark mass values and we 
determined RCs extrapolating their lattice estimators to the chiral limit.  
We employed the RI-MOM scheme~\cite{RIMOM} and the
techniques already used for $N_f=2$ RCs~\cite{Constantinou:2010gr}. However,
for the case of $N_f=4$ simulations with the action and at the lattice
spacing values we are currently using, the implementation of maximal
twist (i.e. tuning the PCAC quark mass to zero), which would guarantee
O($a$) improvement of RCs, is not a trivial task. In fact
in the region of small PCAC quark mass values simulation instabilities
occur that lead for it to very large autocorrelation times.
Hence we opted for an alternative way, already
proposed in Ref.~\cite{FR1}, to achieve 
O$(a)$ improvement though working out of maximal twist.
The method, being based on averaging results obtained at
opposite values of the PCAC quark mass, entails the need
of doubling the reasonably low CPU time cost for producing
gauge simulations at non-zero standard and twisted quark mass.
The present contribution is a report of a work in progress. A first presentation 
and numerical test of our method appeared already in Ref.~\cite{RCs_LAT2010}.

\subsection{Computational setup}
We consider the following fermionic lattice action (written in the so called physical basis)  
\begin{equation}
S_F^{ph}= a^4 \sum_{x}\sum_{f=1}^4 \bar{q}_f \left[ \gamma\cdot\widetilde{\nabla}
-i\gamma_5 r_f e^{i\gamma_5 r_f\theta_{0,f}} (- \dfrac{a}{2} \nabla^*\nabla + m_{\rm cr}) 
+ M_{0,f} \right] q_f(x)  \, ,
\end{equation}
where $q_f$ denotes a singlet quark flavour and $r_f$ takes values either -1 or +1. 
The chiral quark field rotation 
$\chi_f\to q_f = \exp [\frac{i}{2}(\frac{\pi}{2} -\theta_{0,f}) \gamma_5 r_f]\chi_f$
brings the action into the so-called twisted basis~\footnote{For consistency
with the standard Wilson fermion notation, the operator RCs are named according
to the form the operators take in this basis, where the Wilson term is untwisted.}
\beq
S_F^{tm}=a^4 \sum_{x}\sum_{f=1}^4 \bar\chi_f \left[ \gamma\cdot\widetilde{\nabla}
- \dfrac{a}{2} \nabla^*\nabla + m_{0,f} + i\gamma_5 r_f \mu_f \right]\chi_f(x)
\eeq   
The bare mass parameters and the angle $\theta_0$ are given by
\begin{equation}
M_{0,f} = \sqrt{ (m_{0,f} - m_{\rm cr})^2 + \mu_f^2 }\,,\quad
\sin \theta_{0,f} = \frac{m_{0,f} - m_{\rm cr}}{M_{0,f}}\,, \quad
\cos \theta_{0,f} = \frac{\mu_f}{M_{0,f}}\,.
\end{equation}
In practice we make use of $m_{\rm{PCAC}}$ to estimate $(m_{0,f} - m_{\rm cr})$.  
In this way we take as the renormalised quantities the polar quark mass 
$\hat{M}_f = Z_P^{-1} M_f = Z_P^{-1} \sqrt{ Z_A^2 m_{\rm PCAC}^2 + \mu_f^2 }$ and the angle $\theta_f$, 
complementary to the twisted angle $\omega_f$ ($\theta_f = \pi/2 - \omega_f$), given by 
$\tan \theta_f = Z_A m_{\rm PCAC}/\mu_f$. 
As we use four mass degenerate quark flavours and we adopt 
a partially quenched setup, the knowledge of the four parameters 
$M_{\rm sea},\theta_{\rm sea}, M_{\rm val},\theta_{\rm val}$ is sufficient 
to describe our RC computation.

In the following we focus on the evaluation of
the RCs of the (non-singlet) quark bilinear 
operators\footnote{For the 
computation of the RCs of the four-fermion operators 
using the same setup see Ref.~\cite{BK2p1p1}.} 
$O_\Gamma=\bar{\chi}_f \Gamma \chi_{f'}$  where 
$\Gamma\;=\;S, P, V, A, T$, in the RI' variant of the RI-MOM scheme.
One first determines the 
quark field RC, $Z_q$, through
\beq
Z_q^{-1}\frac{-i}{12N(p)}\sum_\rho\mbox{}^\prime\left[ 
\frac{\mathrm{Tr}(\gamma_\rho S_f(p)^{-1})}{\tilde{p}_\rho}\right]_{\tilde{p}^2=\mu^2} 
\,=\,1 \, , \qquad 
\mathrm{any\, }f  \, 
\eeq    
where $\tilde{p}_\mu\equiv\tfrac{1}{a}\sin ap_\mu$, 
$\tilde{p}^2=\sum_\mu\tilde{p}_\mu^2 $ and 
$S_f(p) \,=\,a^4\sum_x\,e^{-ipx}\left\langle
\chi_f(x)\bar{\chi_f}(0)\right\rangle$ 
is the Landau gauge quark propagator in momentum space. 
The sum $\sum_\rho\mbox{}^\prime$ runs over the Lorentz indices for 
which $\tilde p_\rho$ is different from zero and $N(p)=\sum_\rho\mbox{}^\prime 1$.
Then one computes the RC, $Z_\Gamma$, of the operator $O_\Gamma$ 
via
\beq
Z_q^{-1}Z^{(ff^\prime)}_\Gamma\mathrm{Tr}\left[
\left(  S_f^{-1}(\tilde{p})G_\Gamma^{(ff^\prime)}(\tilde{p},\tilde{p})S_{f^\prime}^{-1}(\tilde{p})
 \right) P_\Gamma\right]_{\tilde{p}^2=\mu^2} 
\,=\,1  \, ,  \qquad f\neq f^{\prime} \, .
\label{ZGam} 
\eeq
where
\beq
G^{(ff^\prime)}_\Gamma(p,p)\,=\,a^8\sum_{x,y}\,e^{-ip(x-y)}
\left\langle \chi_f(x)(\bar{\chi}_f\Gamma\chi_{f^\prime})(0)\bar{\chi}_{f^\prime}(y)\right\rangle 
\qquad\Gamma\;=\;S, P, V, A, T \, .
\eeq
with $r_{f^\prime}=-r_f$ for the Wilson parameters of the (valence)
quark flavours $f$ and $f'$.
We note that RCs are blind to the choice of $sign(r_f)$ but 
lattice artefacts in their estimators in general are not.  

In our computation we will exploit the fact that 
the $O(a^{2k+1})$ artefacts occurring in the vacuum expectation
values of (multi)local operators $O$ vanish if we take the $\theta$-average defined by
$\frac{1}{2} \Big[
\langle O \rangle|_{\hat{M},\theta} + \langle O \rangle|_{\hat{M},-\theta} \Big]$. 
The O$(a)$ improvement obtained in this way 
is a consequence of the symmetry\footnote{We denote by ${\cal P}$ 
the parity transformation of the fields and by ${\cal D}_d$ the transformation defined from  
${\cal D}_d q_f(x)=e^{3i\pi/2}q_f(-x)$,  ${\cal D}_d \bar{q}_f(x)=e^{3i\pi/2}\bar{q}_f(-x)$ 
and ${\cal D}_d U_{\mu}=U_{\mu}^{\dagger}(-x-a\hat{\mu})$.} 
${\cal P} \times (\theta_0 \to -\theta_0) \times 
{\cal D}_d \times (M_0 \to - M_0)$ of the lattice action and occurs for 
operator expectation values and form factors that are invariant under 
${\cal P} \times (\theta_0 \to -\theta_0)$, see refs.~\cite{FR1,FMPR}.
In particular this holds for our RCs estimators 
at any value of quark mass, $M_f$, and momentum, $\tilde{p}$.

\section{Analysis and Results}
We have produced $N_f=4$ dynamical quark gauge configurations at three values of 
the inverse gauge coupling, $\beta=1.90$, 
1.95 and 2.10, and for each $\beta$ at a number of values of $M^{\rm{sea}}$ and 
(nearly) opposite values of $\theta^{\rm sea}$ (the ${\tt p/m}$ in the ensemble labels
refers to $sign(\theta_{\rm{sea}})$). An overview of these ensembles and the valence
mass parameters chosen for the Landau gauge correlation functions is given in
Table~\ref{param_details}. 
Some quark propagator computations are still in progress,
as indicated. Hence, the corresponding ensembles can not be used in this analysis.

\begin{table}[!ht]
\begin{center}
\footnotesize
\begin{tabular}{|c|c|c|c|c|c|c|}
\hline
ensemble & $a\mu_{{\rm sea}}$ & $am_{{\rm PCAC}}^{{\rm sea}}$ & $aM_{{\rm 0}}^{{\rm sea}}$ & $\theta^{{\rm sea}}$ & $
a\mu_{{\rm val}}$ & $am_{{\rm PCAC}}^{{\rm val}}$\tabularnewline
\hline
\hline
\multicolumn{7}{|c|}{$\beta=1.90$}\tabularnewline
\hline
${\tt 4m}$ & 0.0080 & -0.0390(01) & 0.0285(01) & -1.286(01) & in progress & ... \tabularnewline
${\tt 4p}$ & 0.0080 & 0.0398(01)  & 0.0290(01) & +1.291(01) & in progress & ... \tabularnewline
\hline
${\tt 3m}$ & 0.0080 & -0.0358(02) & 0.0263(01) & -1.262(02) & in progress & ... \tabularnewline
${\tt 3p}$ & 0.0080 & +0.0356(02) & 0.0262(01) & +1.260(02) & in progress & ...\tabularnewline
\hline
${\tt 2m}$ & 0.0080 & -0.0318(01) & 0.0237(01) & -1.226(02) & in progress  & ... \tabularnewline
${\tt 2p}$ & 0.0080 & +0.0310(02) & 0.0231(01) & +1.218(02) & in progress & ... \tabularnewline
\hline
${\tt 1m}$ & 0.0080 & -0.0273(02) & 0.0207(01) & -1.174(03) & in progress  & ...\tabularnewline
${\tt 1p}$ & 0.0080 & +0.0275(04) & 0.0209(01) & +1.177(05) & in progress  & ... \tabularnewline
\hline
\multicolumn{7}{|c|}{$\beta=1.95$}\tabularnewline
\hline
${\tt 1m}$ & 0.0085 & -0.0413(02) & 0.0329(01) & -1.309(01) & $[0.0085,\ldots,0.0298]$ & -0.0216(02)\tabularnewline
${\tt 1p}$ & 0.0085 & +0.0425(02) & 0.0338(01) & +1.317(01) & $[0.0085,\ldots,0.0298]$ & +0.0195(02)\tabularnewline
\hline
${\tt 7m}$ & 0.0085 & -0.0353(01) & 0.0285(01) & -1.268(01) & $[0.0085, \ldots, 0.0298]$ & -0.0180(02)\tabularnewline
${\tt 7p}$ & 0.0085 & +0.0361(01) & 0.0285(01) & +1.268(01) & $[0.0085, \ldots, 0.0298]$ & +0.0181(01)\tabularnewline
\hline
${\tt 8m}$ & 0.0020 & -0.0363(01) & 0.0280(01) & -1.499(01) & $[0.0085, \ldots, 0.0298]$ & -0.0194(01)\tabularnewline
${\tt 8p}$ & 0.0020 & +0.0363(01) & 0.0274(01) & +1.498(01) & $[0.0085, \ldots, 0.0298]$ & +0.0183(02)\tabularnewline
\hline
${\tt 3m}$ & 0.0180 & -0.0160(02) & 0.0218(01) & -0.601(06) & $[0.0060,\ldots,0.0298]$ & -0.0160(02)\tabularnewline
${\tt 3p}$ & 0.0180 & +0.0163(02) & 0.0219(01) & +0.610(06) & $[0.0060,\ldots,0.0298]$ & +0.0162(02)\tabularnewline
\hline
${\tt 2m}$ & 0.0085 & -0.0209(02) & 0.0182(01) & -1.085(03) & $[0.0085,\ldots,0.0298]$ & -0.0213(02)\tabularnewline
${\tt 2p}$ & 0.0085 & +0.0191(02) & 0.0170(02) & +1.046(06) & $[0.0085,\ldots,0.0298]$ & +0.0191(02)\tabularnewline
\hline
${\tt 4m}$ & 0.0085 & -0.0146(02) & 0.0141(01) & -0.923(04) & $[0.0060,\ldots,0.0298]$ & -0.0146(02)\tabularnewline
${\tt 4p}$ & 0.0085 & +0.0151(02) & 0.0144(01) & +0.940(07) & $[0.0060,\ldots,0.0298]$ & 0.0151(02)\tabularnewline
\hline
\multicolumn{7}{|c|}{$\beta=2.10$}\tabularnewline
\hline 
${\tt 5m}$ & 0.0078 & -0.00821(11) & 0.0102(01) & -0.700(07) & $[ 0.0048, \ldots, 0.0293 ]$ & -0.0082(01)\tabularnewline
${\tt 5p}$ & 0.0078 & 0.00823(08)  & 0.0102(01) & +0.701(05) & $[ 0.0048, \ldots, 0.0293 ]$ & +0.0082(01)\tabularnewline
\hline 
${\tt 4m}$ & 0.0064 & -0.00682(13) & 0.0084(01) & -0.706(09) & in progress & ... \tabularnewline
${\tt 4p}$ & 0.0064 & +0.00685(12) & 0.0084(01) & +0.708(09) & in progress & ... \tabularnewline
\hline 
${\tt 3m}$ & 0.0046 & -0.00585(08) & 0.0066(01) & -0.794(07) &  $[  0.0025, \ldots, 0.0297 ]$ & -0.0059(01)\tabularnewline
${\tt 3p}$ & 0.0046 & +0.00559(14) & 0.0064(01) & +0.771(13) & $[  0.0025, \ldots, 0.0297 ]$ & +0.0056(01)\tabularnewline
\hline
${\tt 2m}$ & 0.0030 & -0.00403(14) & 0.0044(01) & -0.821(17)  & in progress & ...\tabularnewline
${\tt 2p}$ & 0.0030 & +0.00421(13) & 0.0045(01) & +0.843(15)  & in progress & ...\tabularnewline
\hline
\end{tabular}\caption{Overview of produced ensembles at $\beta=1.90$, 1.95  and 2.10}
\label{param_details}
\end{center}
\end{table}

\noindent The basic ingredient of the calculation, due to Eq.(\ref{ZGam}), is the lattice RC estimator, 
$$Z_{\Gamma}^{\tt Np/m} \equiv Z_{\Gamma}^{\tt Np/m}(M_0^{\rm{sea}, {\tt Np/m}}, 
\theta_0^{\rm{sea}, {\tt Np/m}}; 
\{ M_j^{\rm{val}, {\tt Np/m}}, \theta_j^{\rm{val}, {\tt Np/m}} \}; \tilde{p}^2; \beta), $$ 
where $j$ labels the valence quark polar mass and the 
momenta $p$ are such
that $a^2\tilde{p}^2$ ranges from $0.5$ to $~2.5$ and that 
$a^2 \tilde{p}^{[4]}/\tilde{p}^2 \leq 0.28$, 
with $\tilde{p}^{[4]}=\sum_\rho \tilde{p}_\rho^4$.
Our analysis goes through the following steps.\\
(1) Subtract from the RC estimator the O($a^2g^2$) cutoff effects at the chiral point, known
    from Ref.~\cite{Constantinou:2009tr}
(2) Build the O($a$) improved estimator 
$Z_{\Gamma}^{\tt N} \equiv Z_{\Gamma}^{\tt N}(M_0^{\rm{sea}, {\tt N}}, \theta_0^{\rm{sea}, {\tt N}}; 
\{ M_j^{\rm{val}, {\tt N}}, \theta_j^{\rm{val}, {\tt N}} \}; \tilde{p}^2; \beta) = 
\frac{1}{2}\left[ Z_{\Gamma}^{\tt Np} + Z_{\Gamma}^{\tt Nm}  \right]. $ \\
(3) Extrapolate to the chiral limit value, first in the valence and then in the sea sector. \\
(4) Evaluate the RCs at a given renormalization scale after taking care of the residual 
    lattice artefacts according to the methods ``M1'' and ``M2'' 
    (see Ref.~\cite{Constantinou:2010gr} and discussion below). 

In step (2) chiral fit Ans\"atze are inspired to the mass parameter dependence 
expected from continuum
QCD and the Symanzik analysis of lattice artefacts. For the valence chiral extrapolation
we considered as fit functions linear combinations of constant, $M_{j}^{\rm{val}}$,
$( M_{j}^{\rm{val}})^2$, $M_{j}^{\rm{val}}\, cos(\theta_j^{\rm{val}})$, 
and $(M_{j}^{\rm{val}}\, cos(\theta_j^{\rm{val}}))^2$. For the sea chiral limit
linear combinations of $M_{0}^{\rm{sea}}$, $ (M_{0}^{\rm{sea}})^2$ and 
$cos(\theta_j^{\rm{val}})\, ( M_{0}^{\rm{sea}})^2$ were considered. 
In the case of $Z_P$, just before step (2) we remove the Goldstone pole contribution,
which, depending directly on the lattice pseudoscalar meson mass, happens to be somewhat
different for estimators corresponding to opposite $\theta^{val,sea}$-values.
As for step (4), in the first method (``M1''),
after bringing the RC-estimators to a common renormalization scale
($\tilde{p}^2_\mathrm{M1}=1/a^2$), we remove the remaining
$O(a^2\tilde{p}^2)$ discretization errors by a linear fit in
$\tilde{p}^2$. Here the fit range is $1.5 \leq a^2\tilde{p}^2 \leq 2.2$.
The second method (``M2'') consists in simply taking
the value of the RCs estimators at some high momentum point kept fixed in 
physical units at all $\beta$'s. Here we choose $\tilde{p}^2 = 12.0 \pm 0.5$~GeV$^2$.
The two approaches yield RC values differing only by cutoff effects.

\begin{figure}
\centering
\subfigure[]{
\includegraphics[scale=0.60]{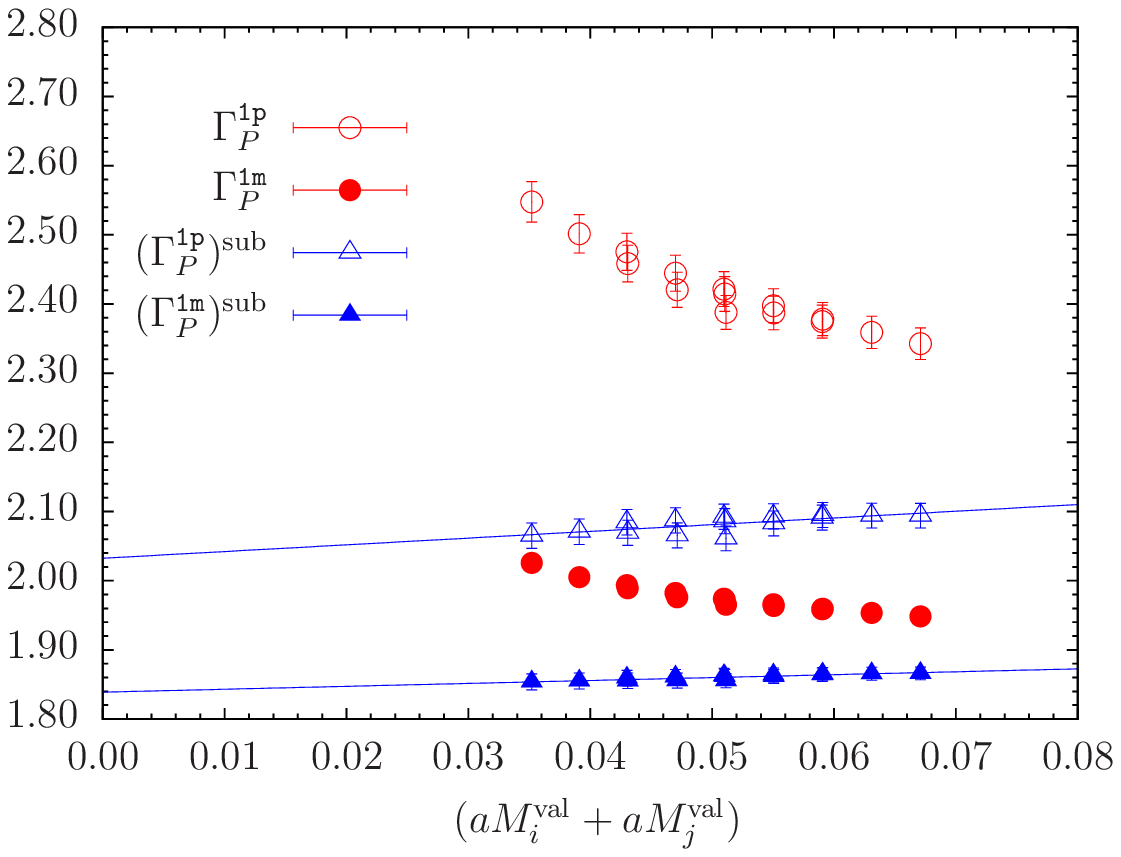}}
\subfigure[]{
\includegraphics[scale=0.60]{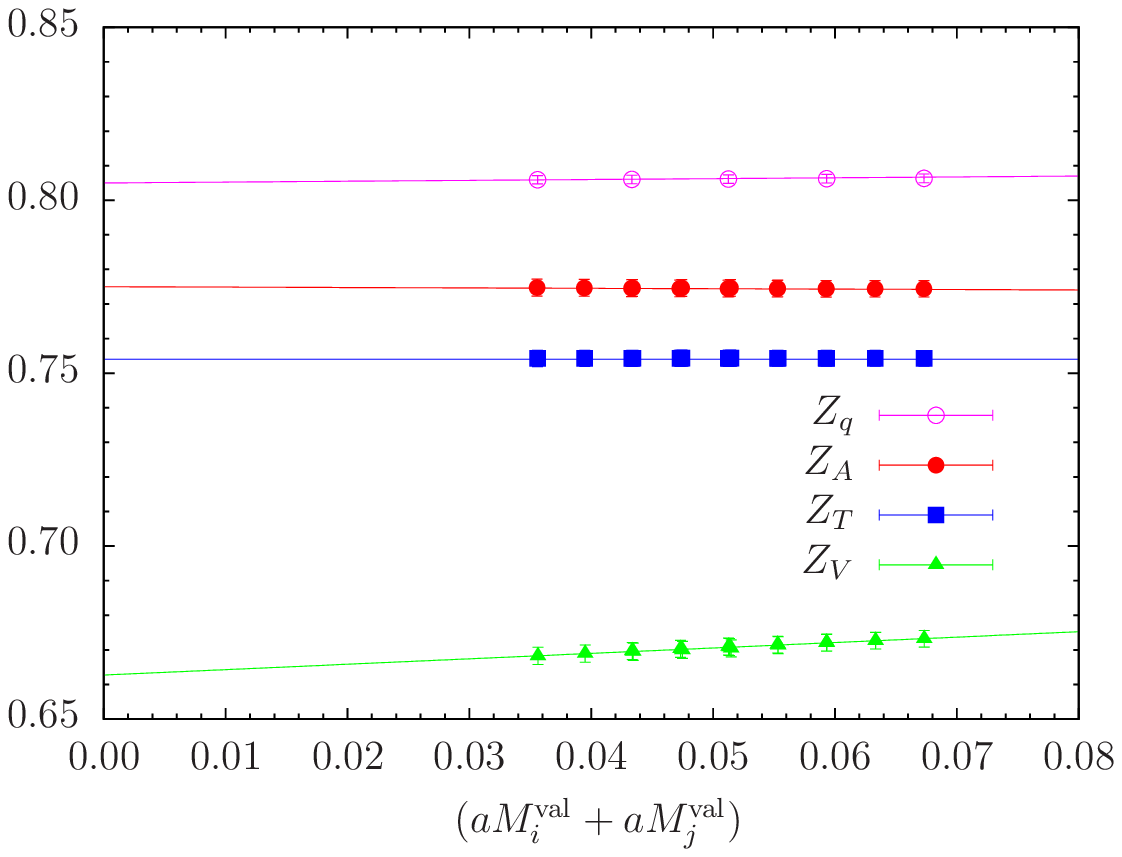}}
\caption{$\beta=1.95$, ensemble ${\tt 1p/m}$: (a) Goldstone pole subtraction fit 
applied separately on ${\tt p}$ and ${\tt m}$ 
for estimators of $Z_P^{-1}$
at $\tilde{p}^2 \sim 9.5\, {\rm GeV}^2$; 
(b) extrapolation of the $\theta$-averaged RC estimators to the chiral limit 
of $Z_q$, $Z_A$, $Z_T$ and $Z_V$  at 
$\tilde{p}^2 \sim 11.5 \, {\rm GeV}^2$.
The situation is similar for other ensembles.}
\label{valchiral}
\end{figure}      
 
\begin{figure}
\centering
\subfigure[]{
\includegraphics[scale=0.60]{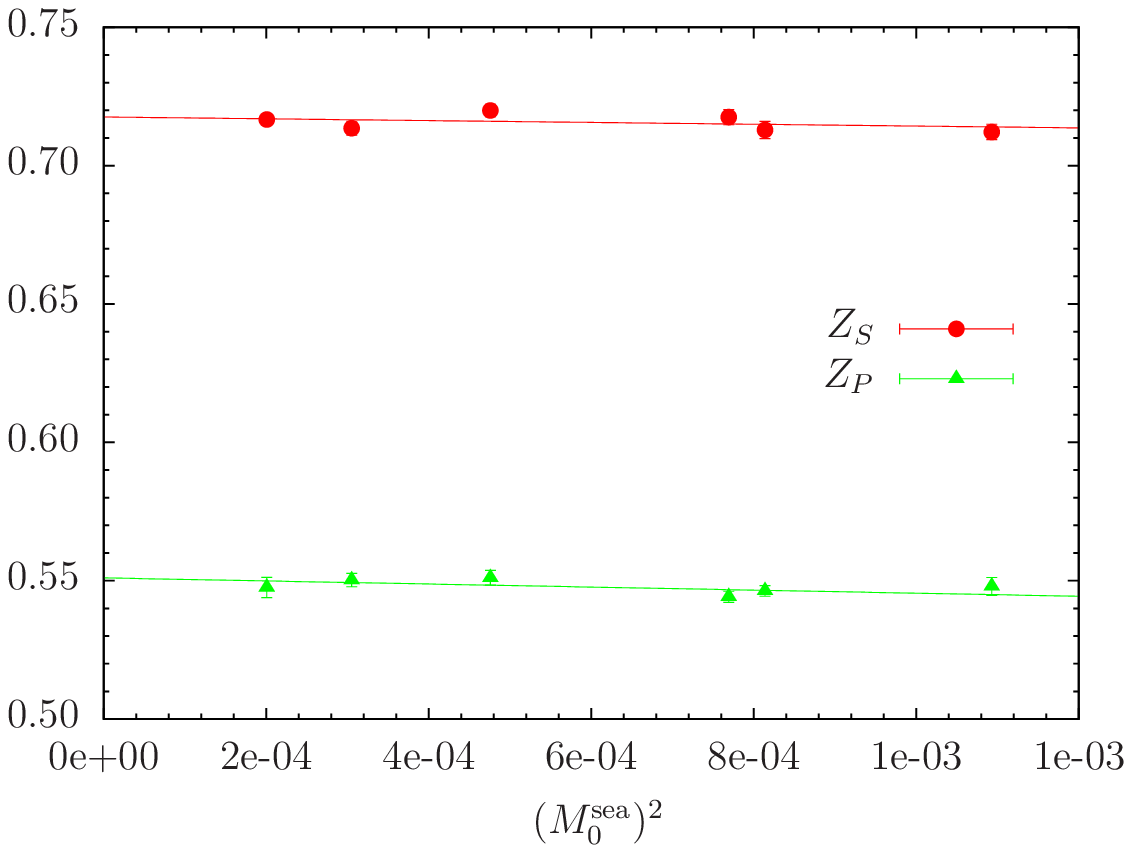}}
\subfigure[]{
\includegraphics[scale=0.60]{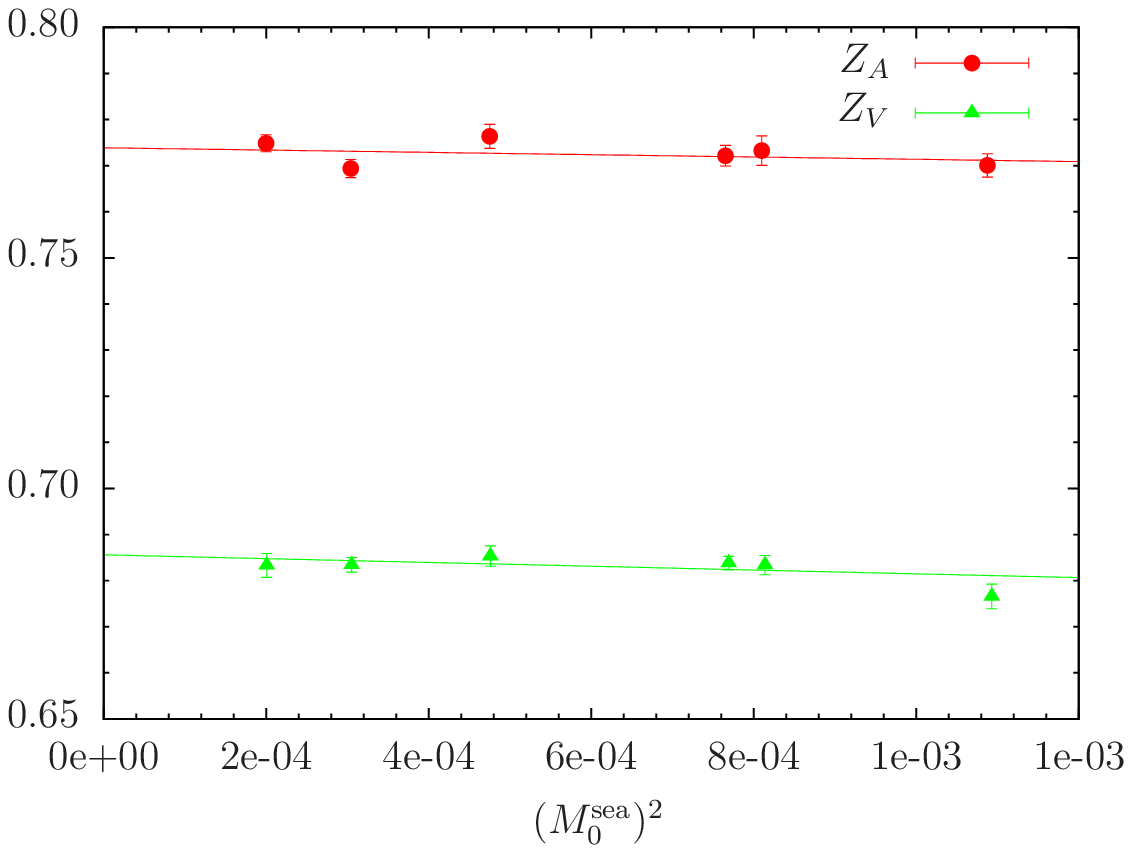}}
\caption{$\beta=1.95$: extrapolation to the sea chiral limit for (a) $Z_P$ and $Z_S$; (b) $Z_A$ and $Z_V$ 
(at $\tilde p^2 \sim 11.5 \, {\rm GeV}^2$).}
\label{seachiral}
\end{figure}

In Fig.~\ref{valchiral}, for the example of the ensemble ${\tt 1p/m}$ of $\beta=1.95$,
we show the Goldstone pole removal and the residual valence quark mass dependence in the
analysis of $Z_P$ (panel (a)) and the extrapolation to the valence chiral limit for 
$Z_q$, $Z_A$, $Z_T$ and $Z_V$ (panel (b)). The fit Ansatz is a linear function 
of the valence quark polar mass. We checked that results do not change significantly
by using more complicated fit functions (involving higher mass powers or $\theta^{\rm val}$). 
Fig.~\ref{seachiral} shows, for $\beta=1.95$ the extrapolation to the sea chiral limit of 
$Z_P$, $Z_S$ (panel (a)) and $Z_A$,  $Z_V$ 
(panel (b)). The 
fit Ansatz is a linear function of 
$(M^{\rm{sea}})^2$. 
More elaborated fit functions give compatible results.

\begin{figure}
\centering
\subfigure[]{
\includegraphics[scale=0.60]{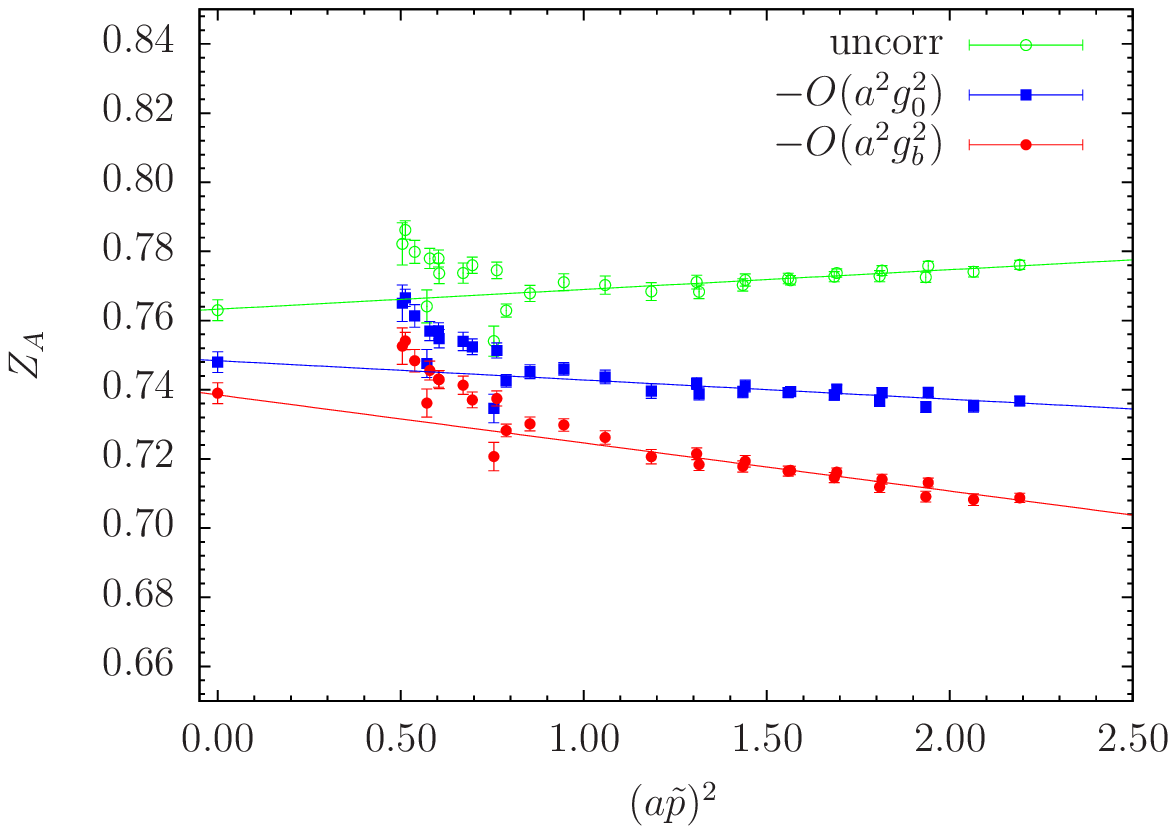}}
\subfigure[]{
\includegraphics[scale=0.60]{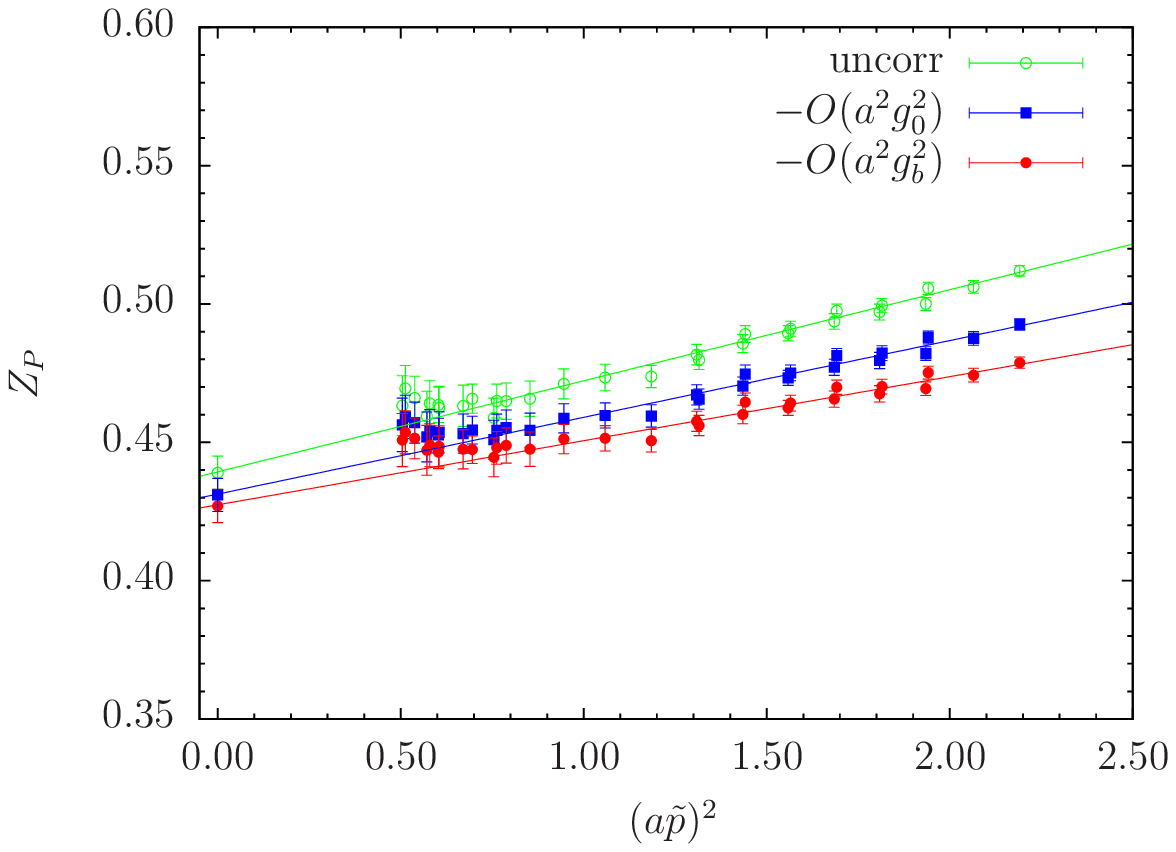}}
\caption{$\beta=1.95$; dependence of $Z_A$ (left panel) and $Z_P$ (right panel) 
on $(a^2\tilde{p}^2)$. Uncorrected and one-loop corrected (with two
choices of the gauge coupling, bare and plaquette-boosted)  
RC estimators are shown. }
\label{ZAZPvsp2_b195}
\end{figure}

\begin{figure}[!h]
\centering
\subfigure[]{
\includegraphics[scale=0.60]{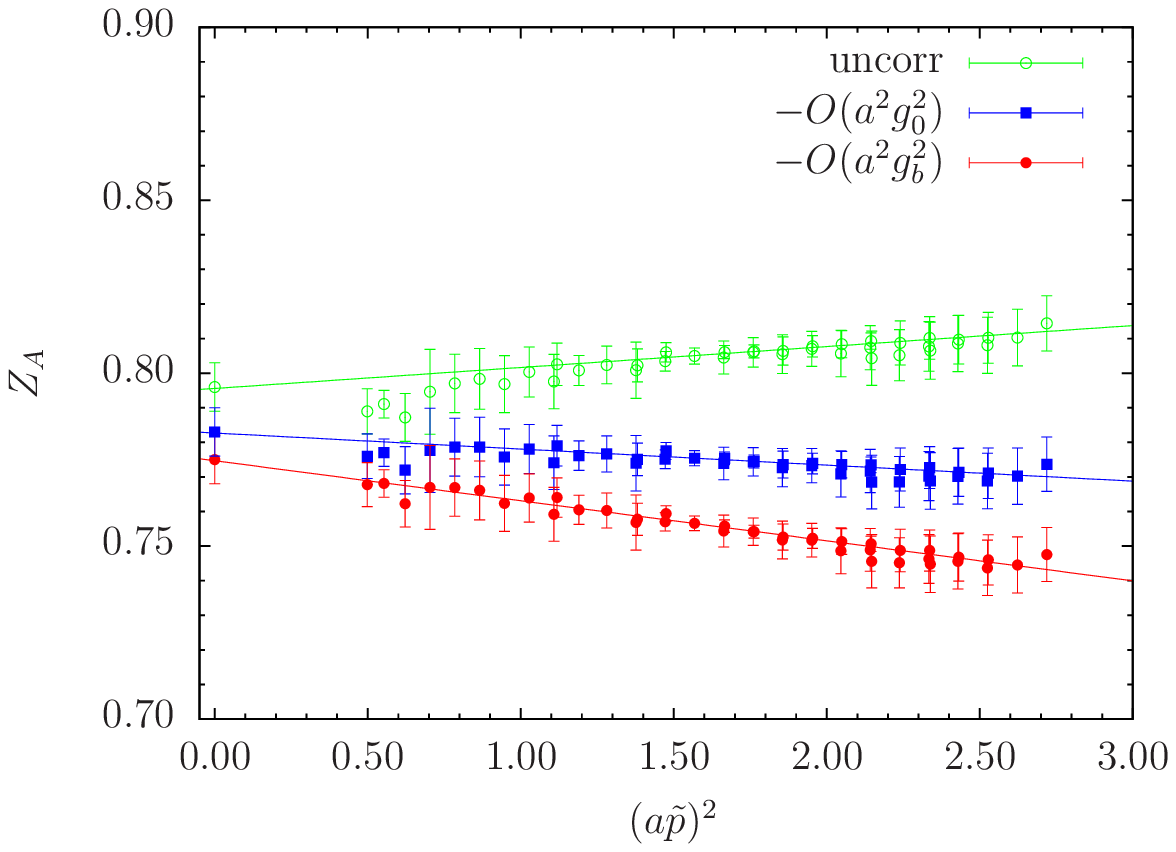}}
\subfigure[]{
\includegraphics[scale=0.60]{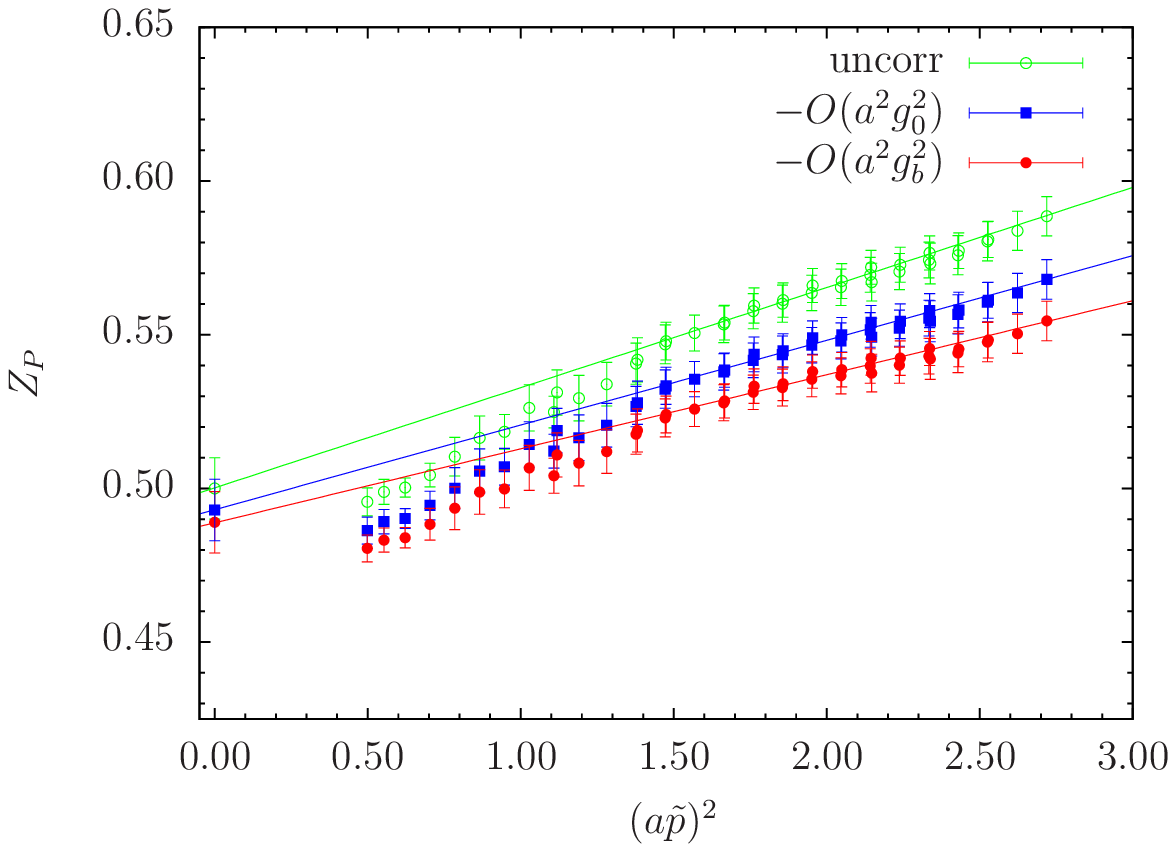}}
\caption{$\beta=2.10$; dependence of $Z_A$ (left panel) and $Z_P$ (right panel) on 
$(a^2\tilde{p}^2)$. Uncorrected and one-loop corrected (with two choices
of the gauge coupling, bare and plaquette-boosted) 
RC estimators are shown. }
\label{ZAZPvsp2_b210}
\end{figure}

For the typical (and important) cases of $Z_A$ and $Z_P$ we show for $\beta=1.95$
(in Fig.~\ref{ZAZPvsp2_b195}) and $\beta=2.1$ (in Fig.~\ref{ZAZPvsp2_b210}) the 
residual dependence on $a^2\tilde{p}^2$ of RC-estimators (in the case of $Z_P$ brought 
to a common renormalization scale 
($1/a(\beta)$) via three-loop evolution). The nice quality of the
linear fit leading to the ``M1'' RC-values is visible, while results of the ``M2''
type are obtained from data at $a^2\tilde{p}^2$ in the range (1.8-2.0) and (1.10-1.23)
for $\beta=1.95$ and $\beta=2.1$, respectively. In each plot three different RC-estimators
are considered, which differ from each other in the way the (beneficial)  
subtraction of O($a^2g^2$) lattice artefacts of step (1) is carried out.

\noindent In Table~\ref{results} we gather our {\em preliminary} results for the RCs 
at two values of the gauge coupling, $\beta=1.95$ and 
$\beta=2.10$. We present results obtained from the two methods described above, namely 
``M1" and ``M2". Perturbative contibutions O$(a^2g^2)$ have been subtracted using the 
coupling constant estimate $g_0^2=6/\beta$. 
Results for the RCs 
whose anomalous dimension is non-zero are given at the scale $1/a$ in the RI/MOM scheme.
Following ref.~\cite{ETMC2p1p1}, we take $a^{-1}|_{\beta=1.95(2.1)}= 2.5(3.2)$ GeV. 

\begin{table}[!h]\label{results}
\begin{center}
\footnotesize
\begin{tabular}{|c|c|c|c|c|c|c|c|}
\hline 
Method & $Z_{A}$ & $Z_{V}$ & $Z_{P}(1/a)$ & $Z_{S}(1/a)$ & $Z_{P}/Z_{S}$ & $Z_{T}(1/a)$ & $Z_{q}(1/a)$\tabularnewline
\hline
\hline 
\multicolumn{8}{|c|}{$\beta=1.95$}\tabularnewline
\hline 
M1 & 0.746(05) & 0.614(03) & 0.426(06) & 0.609(08) & 0.700(08) & 0.734(04) & 0.752(05) \tabularnewline
M2 & 0.738(01) & 0.639(02) & 0.483(02) & 0.684(01) & 0.706(03) & 0.734(01) & 0.769(01)\tabularnewline
\hline 
\multicolumn{8}{|c|}{$\beta=2.10$}\tabularnewline
\hline 
M1 & 0.783(07) & 0.683(13) & 0.493(10) & 0.669(08) & 0.737(14) & 0.775(11) & 0.786(13)\tabularnewline
M2 & 0.777(05) & 0.680(05) & 0.515(06) & 0.696(08) & 0.740(08) & 0.771(05) & 0.794(07) \tabularnewline
\hline
\end{tabular}
\end{center}\caption{Our {\em preliminary} results for quark bilinear RCs at $\beta=1.95$ and $\beta=2.10$.}
\end{table}
\vspace*{0cm}

\section{Acknowledgements} \noindent We thank AuroraScience, Donald Smits center for information technology, 
University of 
Groningen,  IDRIS,  INFN/apeNEXT and J\"ulich Supercomputing Center for 
providing the necessary CPU time. We acknowledge PRACE Research Infrastructure
resource based in Germany at Forschungzentrum Juelich (FZJ) under the project PRA027, 
ISCRA at CINECA (Italy) under the application HP10A7IBG7 and
GENCI Grant 052271 and CC-IN2P3 for partial computer support.


\begin{thebibliography}{99}
\bibitem{ETMC2p1p1}
  {\bf ETMC}, R.~Baron {\it et al.},
  JHEP {\bf 1006 } (2010)  111 [1004.5284 [hep-lat]]; \\ 
  \textbf{ETMC}, R.~Baron {\it et al.} 
   Comput.\ Phys.\ Commun.\  {\bf 182 } (2011)  299-316  [1005.2042 [hep-lat]]. 
 \bibitem{Iwasaki}
  Y.~Iwasaki,
  Nucl.\ Phys.\  B {\bf 258} (1985) 141.
\bibitem{FR1}
  R.~Frezzotti and G.~C.~Rossi,
  JHEP {\bf 0408} (2004) 007
  [hep-lat/0306014]
\bibitem{FRnondegproc}
  R.~Frezzotti and G.~C.~Rossi,
  Nucl.\ Phys.\ Proc.\ Suppl.\  {\bf 128} (2004) 193
  [hep-lat/0311008]
\bibitem{FR2}
  R.~Frezzotti and G.~C.~Rossi,
  JHEP\ {\bf 0410} (2004) 070
  [hep-lat/0407002].

\bibitem{RIMOM}
  G.~Martinelli {\it et al.},
  Nucl.\ Phys.\  B {\bf 445} (1995) 81
  [hep-lat/9411010].
\bibitem{Constantinou:2010gr}
  {\bf ETMC}, M.~Constantinou {\it et al.}  ,
  JHEP {\bf 1008} (2010) 068
  [1004.1115 [hep-lat]].
\bibitem{RCs_LAT2010}
  {\bf ETMC}, P.~Dimopoulos {\it et al.},
  PoS {\bf LATTICE2010 } (2010)  235.
  [1101.1877 [hep-lat]].  
\bibitem{FMPR}
  R.~Frezzotti, G.~Martinelli, M.~Papinutto, G.~C.~Rossi,
  JHEP {\bf 0604 } (2006)  038.
  [hep-lat/0503034].  
\bibitem{BK2p1p1}
 {\bf ETMC}, N.~Carrasco {\it et al.}, 
  PoS \textbf{Lattice 2011} (2011) 276 
  [1111.1262 [hep-lat]].
\bibitem{Constantinou:2009tr}
  M.~Constantinou {\it et al.}
  JHEP {\bf 0910} (2009) 064
  [0907.0381 [hep-lat]].
\end{thebibliography}
\end{document}